\documentclass[aps,pra,reprint,showkeys,superscriptaddress]{revtex4-1}
\usepackage{graphicx}
\usepackage{dcolumn}
\usepackage{bm}
\usepackage{amsmath}
\allowdisplaybreaks[4]

\begin{document}



\def\a{\alpha}
\def\b{\beta}
\def\d{\delta}
\def\e{\epsilon}
\def\g{\gamma}
\def\h{\mathfrak{h}}
\def\k{\kappa}
\def\l{\lambda}
\def\o{\omega}
\def\p{\wp}
\def\r{\rho}
\def\t{\tau}
\def\s{\sigma}
\def\z{\zeta}
\def\x{\xi}
\def\V={{{\bf\rm{V}}}}
 \def\A{{\cal{A}}}
 \def\B{{\cal{B}}}
 \def\C{{\cal{C}}}
 \def\D{{\cal{D}}}
\def\K{{\cal{K}}}
\def\O{\Omega}
\def\R{\bar{R}}
\def\T{{\cal{T}}}
\def\L{\Lambda}
\def\f{E_{\tau,\eta}(sl_2)}
\def\E{E_{\tau,\eta}(sl_n)}
\def\Zb{\mathbb{Z}}
\def\Cb{\mathbb{C}}

\def\R{\overline{R}}

\def\beq{\begin{equation}}
\def\eeq{\end{equation}}
\def\bea{\begin{eqnarray}}
\def\eea{\end{eqnarray}}
\def\ba{\begin{array}}
\def\ea{\end{array}}
\def\no{\nonumber}
\def\le{\langle}
\def\re{\rangle}
\def\lt{\left}
\def\rt{\right}

\preprint{APS/123-QED}



\title{Exact solution of an integrable anisotropic $J_1-J_2$ spin chain model}

\author{Yi Qiao}
\affiliation{Institute of Modern Physics, Northwest University, Xian 710127, China}
\affiliation{Beijing National Laboratory for Condensed Matter Physics, Institute of Physics, Chinese Academy of Sciences, Beijing 100190, China}
\affiliation{Shaanxi Key Laboratory for Theoretical Physics Frontiers,  Xian 710127, China}

\author{Pei Sun}
\affiliation{Institute of Modern Physics, Northwest University, Xian 710127, China}
\affiliation{Shaanxi Key Laboratory for Theoretical Physics Frontiers,  Xian 710127, China}

\author{Zhirong Xin}
\affiliation{Institute of Modern Physics, Northwest University, Xian 710127, China}
\affiliation{Shaanxi Key Laboratory for Theoretical Physics Frontiers,  Xian 710127, China}

\author{Junpeng Cao}
\email{Corresponding author: junpengcao@iphy.ac.cn}
\affiliation{Beijing National Laboratory for Condensed Matter Physics, Institute of Physics, Chinese Academy of Sciences, Beijing 100190, China}
\affiliation{School of Physical Sciences, University of Chinese Academy of Sciences, Beijing, China}
\affiliation{Songshan Lake Materials Laboratory, Dongguan, Guangdong 523808, China}

\author{Wen-Li Yang}
\email{Corresponding author: wlyang@nwu.edu.cn}
\affiliation{Institute of Modern Physics, Northwest University, Xian 710127, China}
\affiliation{Shaanxi Key Laboratory for Theoretical Physics Frontiers,  Xian 710127, China}
\affiliation{School of Physics, Northwest University, Xian 710127, China}

\date{\today}

\begin{abstract}
An integrable anisotropic Heisenberg spin chain with nearest-neighbour couplings, next-nearest-neighbour couplings and scalar chirality terms is constructed. After proving the integrability, we obtain the exact solution of the system. The ground state and the elementary excitations are also studied. It is shown that the spinon excitation of the present model possesses a novel triple arched structure.
The elementary excitation is gapless if the anisotropic parameter $\eta$ is real
while the elementary excitation has an enhanced gap by the
next-nearest-neighbour and chiral three-spin interactions if the anisotropic parameter $\eta$ is imaginary.
The method of this paper provides a general way to construct new integrable models with next-nearest-neighbour interactions.

\end{abstract}

\keywords{Quantum spin chain; Bethe Ansatz; Yang-Baxter equation}

\maketitle


\section{Introduction}

It is well known that the Heisenberg model has played an important role to account for magnetism in condensed matters. An interesting fact is that this model in one-dimension is exactly solvable \cite{Bet31}. Based on Bethe's exact solution, the ground state energy \cite{Hul38}, the low-lying excitation spectrum \cite{des62} and the magnetization at zero temperature \cite{Gri64} had been studied extensively. This exact solution provided a benchmark to understand a variety of physical phenomena in low-dimensions such as the Luttinger liquid behavior and the fractional excitations. In addition, this model also becomes a typical model in developing new theoretical methods to approach general quantum integrable systems \cite{Tak791,Kor93,Tak99,Wan15}.

Besides the Heisenberg model with nearest-neighbor (NN) exchanges, its generalization with next-nearest-neighbor (NNN) interactions, known as the $J_1-J_2$ model, also attracted a lot of attentions \cite{93,97,98,Sha81,Nom92,Jaf06,Djo16}. The $J_1-J_2$ model is interesting because there exists a topological phase transition at the point of $J_2/J_1=0.241$ \cite{95,Jaf07}.
At the Majumdar-Ghosh point, $J_2/J_1=0.5$, the model Hamiltonian degenerates into a projector operator and the ground state can be obtained exactly \cite{92}.
The ground state is two-fold degenerated and can be expressed by the direct product of spin singlets, supposed the number of site of the system is even.
Another interesting development is that Frahm proposed an integrable $J_1-J_2$ model containing chiral three-spin interactions \cite{Fra92}. The extra scalar chirality terms are introduced for ensuring the integrability. Later, Frahm and R\"{o}denbeck studied the properties of the chiral spin liquid state in the system \cite{Fra97}. Wen, Wilczek and Zee \cite{Wen89} and Baskaran \cite{Bas89} proposed that the expectation value of the spin chirality operator¡± can be used as the order parameter for chiral spin liquids \cite{Kal87}.
Recently, the models with chirality terms have attracted renewed interest in the context of quantum spin liquids \cite{Pop93,Gor15,Che17}.

In this paper, we propose a systemic method to construct new integrable models with the NNN and the scalar chirality terms interactions.
We use the anisotropic XXZ quantum spin chain as an example to show the validity of the method. Before that, we first introduce the main result that we construct an
integrable anisotropic $J_1-J_2$ spin chain with the Hamiltonian
\bea\label{Ham}
\!\!\! H &=&\sum^{2N}_{j=1} \bigg\{ \cos(2a)(\sigma_j^x \sigma_{j+1}^x+\sigma_j^y \sigma_{j+1}^y)+\cos\eta \sigma_j^z \sigma_{j+1}^z \no \\
&&\hspace{-0.8cm} -\frac{\sin^2(2a)\cos\eta}{2\sin^2\eta}\vec{\sigma}_j \cdot \vec{\sigma}_{j+2}+\frac{(-1)^j i \sin(2a)}{2 \sin\eta} \big\{ \cos\eta \vec{\sigma}_{j+1} \!\cdot\!(\vec{\sigma}_{j}  \no \\
&&\hspace{-0.8cm} \times \vec{\sigma}_{j+2} )+[\cos(2a)\!-\!\cos\eta] \sigma_{j+1}^z(\sigma_{j}^x\sigma_{j+2}^y \!-\! \sigma_{j}^y\sigma_{j+2}^x) \big\}\! \bigg\},
\eea
where $\vec{\sigma}_j\equiv (\sigma^x_j,\, \sigma^y_j,\, \sigma^z_j)$ are the Pauli matrices at site $j$, $a$ and $\eta$ are the generic constants describing the coupling strengths,
and the periodic boundary condition
\bea
\vec{\sigma}_{2N+j}=\vec{\sigma}_{j},\quad  j=1,\cdots, 2N,\label{Periodic-C}
\eea
is imposed. The first two terms describe an anisotropic NN interaction, the third term is an isotropic  NNN interaction (i.e., the J2 term) and the last  one corresponds to an
anisotropic  chiral three-spin interaction. We shall show that the anisotropic $J_1-J_2$ spin chain with the Hamiltonian (\ref{Ham}) is integrable and can be exactly solved by the
Bethe ansatz.

Some remarks are in order. (i) The hermitian of the Hamiltonian (\ref{Ham}) requires that $a$ must be real if $\eta$ is imaginary (gapped regime),
and $a$ must be imaginary if $\eta$ is real (gapless regime). (ii) The NN interactions are anisotropic while the NNN interactions are isotropic.
(iii) The anisotropic scalar chirality terms are added to ensure the integrability of the system.
(iv) The coupling strengths in the NNN terms and those in the scalar chirality terms are not independent
but related by the parameters $a$ and $\eta$. (v) The model degenerates into the conventional XXZ spin chain at the points of $a=n\pi$ with integer $n$. (vi) After parameterizing
$a=\bar{a}\,\epsilon$, $\eta=\epsilon$ and then taking the limit of $\epsilon\rightarrow 0$,  our Hamiltonian (\ref{Ham}) becomes
\bea
H &=&\sum^{2N}_{j=1}\left\{ \vec{\sigma}_j \cdot \vec{\sigma}_{j+1}-2\bar{a}^2\, \vec{\sigma}_j \cdot \vec{\sigma}_{j+2}\right.\no\\
&&\quad\quad+(-1)^ji\bar{a}\, \left.\vec{\sigma}_{j+1} \cdot ( \vec{\sigma}_j \times
 \vec{\sigma}_{j+2} )\right\}.\label{Ham-iso}
\eea
The resulting Hamiltonian describe an integrable isotropic $J_1-J_2$ spin chain model, which was studied  previously by Frahm et al \cite{Fra92, Fra97}.

The paper is organized as follows. The model is constructed and the integrability is proved
in section \ref{2}. The exact energy spectrum and the Bethe Ansatz equations are derived in section \ref{3}.
The ground state energy and spinon elementary excitation for real $\eta$ are given in section \ref{4} and
the corresponding results with imaginary $\eta$ are given in section \ref{5}. The results of the non-hermitian case are discussed in section \ref{6} and
section \ref{7} is attributed to the concluding remarks.

\section{Model and Integrability}\label{2}

Throughout, ${V}$ denotes a two-dimensional linear space and let $\{|m\rangle, m=0,1\}$ be an orthogonal basis of it.
We shall adopt the standard notations: for any matrix $A\in {\rm End}({ V})$, $A_j$ is an
embedding operator in the tensor space ${ V}\otimes
{ V}\otimes\cdots$, which acts as $A$ on the $j$-th space and as
identity on the other factor spaces. For $B\in {\rm End}({ V}\otimes { V})$, $B_{ij}$ is an embedding
operator of $B$ in the tensor space, which acts as identity
on the factor spaces except for the $i$-th and $j$-th ones.

Let us introduce the $R$-matrix $R_{0,j}(u)\in {\rm End}({ V}_0\otimes { V}_j)$
\begin{eqnarray}\label{R-matrix}
R_{0,j}(u)\!&=&\! \frac{1}{2}\! \bigg[ \frac{\sin(u\!+\!\eta)}{\sin \eta} (1\!+\!\sigma^z_0 \sigma^z_j) \!+\! \frac{\sin u}{\sin \eta} (1\!-\!\sigma^z_0 \sigma^z_j)
 \bigg] \no \\ && + \frac{1}{2} (\sigma^x_0 \sigma^x_j +\sigma^y_0 \sigma^y_j)
\no \\ &&\hspace{-1cm} \!= \!\frac{1}{\sin\eta} \! \left(\! \begin{array}{cccc}
\sin(u+\eta) & 0    & 0    & 0 \\
0      & \sin u    & \sin\eta & 0 \\
0      & \sin\eta & \sin u    & 0 \\
0      & 0    & 0    & \sin(u+\eta)
\end{array} \!\right)\!,
\end{eqnarray}
where $u$ is the spectral parameter.
The $R$-matrix (\ref{R-matrix}) satisfies the following relations
\bea
&&\hspace{-0.5cm}\mbox{ Initial
condition}:\,R_{0,j}(0)=  P_{0,j},\no \\
&&\hspace{-0.5cm}\mbox{ Unitary
relation}:\,R_{0,j}(u)R_{j,0}(-u)= \phi(u)\times {\rm id},\no \\
&&\hspace{-0.5cm}\mbox{ Crossing
relation}:\,R_{0,j}(u)=-\sigma_0^y R_{0,j}^{t_0}(-u-\eta)\sigma_0^y,\no \\
&&\hspace{-0.5cm}\mbox{ PT-symmetry}:\,R_{0,j}(u)=R_{j,0}(u)=R_{0,j}^{t_0\,t_j}(u),\label{PT}
\eea
where $\phi(u)=-\sin(u+\eta)\sin(u-\eta)/\sin^2\eta$, $t_0$ (or $t_j$) denotes the
transposition in the space ${V}_0$ (or ${V}_j$) and
$P_{0,j}$ is the permutation operator possessing the property
\begin{eqnarray}
  R_{j,k}(u)=P_{0,j}R_{0,k}(u)P_{0,j}.
\end{eqnarray}
The $R$-matrix satisfies the Yang-Baxter equation (YBE)
\bea
&&R_{1,2}(u_1-u_2)R_{1,3}(u_1-u_3)R_{2,3}(u_2-u_3)
\no \\ && \quad =R_{2,3}(u_2-u_3)R_{1,3}(u_1-u_3)R_{1,2}(u_1-u_2).\label{QYB}
\eea

We define the monodromy matrices \cite{Skl88,Wan15} as
\bea\label{monodromy-matrix}
T_0(u)&=&R_{0,1}(u+a) R_{0,2}(u-a) \cdots R_{0,2N-1}(u+a) \no \\ && \times R_{0,2N}(u-a), \no \\
\hat{T}_0(u)&=&R_{0,2N}(u+a) R_{0,2N-1}(u-a) \cdots R_{0,2}(u+a) \no \\ && \times R_{0,1}(u-a),
\eea
where $V_0$ is the auxiliary space, $V_1\otimes V_2 \otimes \cdots \otimes V_{2N}$ is the physical or quantum space,
$2N$ is the number of sites and $a$ is the inhomogeneous parameter.
From the YBE (\ref{QYB}), one can prove that the monodromy matrix $T(u)$ satisfies the Yang-Baxter relation
\bea
R_{1,2}(u-v)T_1(u)T_2(v)= T_2(v)T_1(u)R_{1,2}(u-v).\label{RTT}
\eea
The transfer matrices are the trace of monodromy matrices in the auxiliary space
\bea \label{trans}
t(u)=tr_0 T_0(u), \quad  \hat{t}(u)=tr_0 \hat{T}_0(u).
\eea
Using the crossing symmetry in Eq.(\ref{PT}), we obtain the relations between transfer matrices $t(u)$ and $\hat t(u)$
\bea\label{tt}
t(u)=\hat{t}(-u-\eta), \quad \hat{t}(u)=t(-u-\eta).
\eea
From the Yang-Baxter relation (\ref{RTT}) and Eq.(\ref{tt}), one can prove that the transfer matrices
$t(u)$ [or $\hat t(u)$] with different spectral parameters
commute with each other. Meanwhile, the transfer matrices $t(u)$ and $\hat t(u)$ also commute with each other
\bea
[t(u), t(v)]=[\hat t(u), \hat t(v)] =[t(u), \hat t(u)]=0. \label{t-commu1} \eea
Therefore, both $t(u)$ and $\hat t(u)$ serve as the generating functions of all the conserved quantities of the
system. We note that the transfer matrices $t(u)$ and $\hat t(u)$ can be diagonalized simultaneously.

The model Hamiltonian (\ref{Ham}) can be constructed as (for details, see Appendix A)
\bea\label{Hdef}
H &=& -\frac{N\cos\eta[\cos^2(2a)-\cos(2\eta)]}{\sin^2\eta} +\phi^{1-N}(2a)\sin\eta
\no \\ && \times \bigg\{ \hat{t}(-a)\frac{\partial \, t(u)}{\partial u}\big|_{u=a}+ \hat{t}(a) \frac{\partial \, t(u)}{\partial u}\big|_{u=-a} \bigg\}. \label{t-c2ommu1}
\eea
From the construction (\ref{t-c2ommu1}) and the commutation relation (\ref{t-commu1}) of generating functions $t(u)$ and $\hat t(u)$, we conclude that
the quantum spin chain (\ref{Ham}) with the periodic boundary condition is integrable.

\section{Exact solution}\label{3}

Based on the integrability discussed in the previous section, the Hamiltonian (\ref{Ham}) can be solved exactly via the algebraic Bethe Ansatz \cite{Kor93}.
The matrix form of monodromy matrix $T_0(u)$ in the auxiliary space is
\begin{equation}\label{monodromy-matrix1}
  T_0(u)=\left(
             \begin{array}{cc}
               A(u) & B(u)\\
               C(u) & D(u)\\
             \end{array}
           \right).
\end{equation}
where $A(u)$, $B(u)$, $C(u)$ and $D(u)$ are the operators acting in the quantum space.
We denote the all spins aligning up state as the vacuum state $ |0\rangle$
\bea
|0\rangle=\left(\begin{array}{c}
    1 \\
    0 \\
  \end{array}\right)_1 \otimes\cdots\otimes\left(\begin{array}{c}
    1 \\
    0 \\
  \end{array}\right)_{2N}.\label{left-vacuum}
\eea
The matrix elements of the monodromy matrix $T_0(u)$ acting on the vacuum state gives
\bea
&& A(u)|0\rangle=a(u)|0\rangle, \quad B(u)|0\rangle\neq 0,
\no \\ && C(u)|0\rangle=0, \quad D(u)|0\rangle=d(u)|0\rangle,\label{right-action-1}
\eea
where
\bea\label{adfun}
&&a(u)=\frac{\sin^N(u+a+\eta)\sin^N(u-a+\eta)}{\sin^{2N}\eta}, \no \\
&&d(u)=\frac{\sin^N(u+a)\sin^N(u-a)}{\sin^{2N}\eta}.\no
\eea

From Eq.(\ref{right-action-1}), we know that the operator $B(u)$ can be regarded as the creation operator of all the eigenstates of the system.
Assume the eigenstates take the form
\bea
|\lambda_1,\cdots,\lambda_M\rangle=\prod_{j=1}^{M}B(\lambda_j)|0\rangle, \label{22}
\eea
where $M$ is the number of flipped spins and $\{ \lambda_j \}$ are the Bethe roots.
From the Yang-Baxter relation (\ref{RTT}), we obtain the commutative relations among the elements of monodromy matrix as
\begin{eqnarray}\label{commu-rela}
&&\!\![  A(u)\!,\!  A(v)]\!=\![  B(u)\!,\!  B(v)]\!=\![  C(u)\!,\!  C(v)] \!=\![  D(u)\!,\!  D(v)]\!=\!0,\! \no \\[4pt]
&&\!\!   A(u)  B(v)\!=\!\frac{\sin(u\!-\!v\!-\!\eta)}{\sin(u\!-\!v)}  B(v)
  A(u)\!+\!\frac{\sin\eta}{\sin(u\!-\!v)}  B(u)  A(v),\! \no \\[4pt]
&&\!\!   D(u)  B(v)\!=\!\frac{\sin(u\!-\!v\!+\!\eta)}{\sin(u\!-\!v)}  B(v)  D(u)
\!-\!\frac{\sin\eta}{\sin(u\!-\!v)}  B(u)  D(v),\! \no \\[4pt]
&&\!\![  B(u)\!,\!  C(v)]\!=\!\frac{\sin\eta}{\sin(u\!-\!v)}[  D(v)  A(u)\!-\!  D(u)  A(v)]. \label{RLL-4}
\end{eqnarray}

From the definition (\ref{trans}), the transfer matrix $t(u)$ is
\bea\label{trans1}
t(u)=A(u)+D(u).
\eea
Acting the transfer matrix $t(u)$ on the Bethe state (\ref{22}) and with the help of the commutation relations (\ref{RLL-4}), we have
\bea
t(u)|\lambda_1,\cdots,\lambda_M\rangle &=&\Lambda(u)|\lambda_1,\cdots,\lambda_M\rangle \no \\
&&\hspace{-2.7truecm}+\sum_{j=1}^{M}\Lambda_j(u)B(\lambda_1)\cdots,B(\lambda_{j-1})B(u)B(\lambda_{j+1})\cdots \no \\ &&\hspace{-2.2truecm} \times B(\lambda_M)|0\rangle,\label{wwLam}
\eea
where
\bea
&&\hspace{-0.12truecm}\!\!\Lambda(u)\!=\!a(u)\!\prod_{j=1}^{M}\!\frac{\sin(u\!-\!\lambda_j\!-\!\eta)}{\sin(u-\lambda_j)} \!+\! d(u)\!\prod_{j=1}^{M}\!\frac{\sin(u\!-\!\lambda_j\!+\!\eta)}{\sin(u-\lambda_j)}\!,\label{Lam}  \\
&&\hspace{-0.12truecm}\!\!\Lambda_j(u)=\frac{\sin\eta}{\sin(u-\lambda_j)}\bigg\{a(\lambda_j)\prod_{l\neq j}^{M}\frac{\sin(\lambda_j-\lambda_l-\eta)}{\sin(\lambda_j-\lambda_l)} - d(\lambda_j)
\no \\ && \quad \quad \quad \times\prod_{l\neq j}^{M}\frac{\sin(\lambda_j-\lambda_l+\eta)}{\sin(\lambda_j-\lambda_l)}\bigg\}. \no
\eea
The first term in Eq.(\ref{wwLam}) corresponds to the eigenvalue term, while the last terms in Eq.(\ref{wwLam}) are the unwanted ones.
The  state (\ref{22}) becomes an eigenstate (or the Bethe state) of the transfer matrix provided that the parameters $\{\lambda_j|j=1,\cdots,M\}$ satisfy the Bethe Ansatz equations (BAEs)
\bea
&&\!\! \left[\frac{\sin(\lambda_j\!+\!a\!+\!\eta)\sin(\lambda_j\!-\!a\!+\!\eta)}{\sin(\lambda_j+a) \sin(\lambda_j-a)}\right]^N\!\!\!=\!\prod_{l\neq j}^{M}\frac{\sin(\lambda_j-\lambda_l+\eta)} {\sin(\lambda_j-\lambda_l-\eta)}, \no \\ && \quad \quad j=1,\cdots,M.\label{orBAEs1}
\eea
For convenience, we put $\lambda_j = i u_j/2 - \eta/2$ and $a=ib$ for real $\eta$ and $\lambda_j = u_j/2 - \eta/2$ for imaginary $\eta=i\gamma$ . The BAEs become
\bea\label{orBAEs_real}
&&\left[\frac{\sinh\frac{1}{2}(u_j-2b-i\eta)\sinh\frac{1}{2}(u_j+2b-i\eta)}
{\sinh\frac{1}{2}(u_j-2b+i\eta)\sinh\frac{1}{2}(u_j+2b+i\eta)}\right]^N
\no \\ && \quad =\prod^{M}_{l\neq j}\frac{\sinh\frac{1}{2}(u_j-u_l-2\eta i)}{\sinh\frac{1}{2}(u_j-u_l+2\eta i)}, \quad j=1,\cdots,M.
\eea
for real $\eta$, and
\bea\label{orBAEs_image}
&&\left[\frac{\sin\frac{1}{2}(u_j-2a-i\gamma)\sin\frac{1}{2}(u_j+2a-i\gamma)}
{\sin\frac{1}{2}(u_j-2a+i\gamma)\sin\frac{1}{2}(u_j+2a+i\gamma)}\right]^N
\no \\ && \quad =\prod^{M}_{l\neq j}\frac{\sin\frac{1}{2}(u_j-u_l-2\gamma i)}{\sin\frac{1}{2}(u_j-u_l+2\gamma i)}, \quad j=1,\cdots,M.
\eea
for imaginary $\eta=i\gamma$.
From Eqs.(\ref{tt}), (\ref{Hdef}) and (\ref{Lam}) we obtain the eigenvalue of the Hamiltonian (\ref{Ham}) in terms of the Bethe roots as
\bea\label{energy_real}
&&\hspace{-0.42truecm}E
\!=\!\frac{N\!\cos\eta[\cosh^2(2b)\!-\!\cos(2\eta)]}{\sin^2\eta}\!-\! [\cosh(4 b)\!-\!\cos(2\eta)]
\no \\
&&\hspace{-0.12truecm}\!\times\!\! \sum^{M}_{j=1}\!\bigg\{\! \frac{1}{\cosh(u_j\!+\!2b)\!-\!\cos\eta}\!+\!\frac{1}{\cosh(u_j\!-\!2b)\!-\!\cos\eta} \!\bigg\}\!,
\eea
where $\eta$ is real and $\{u_j\}$ should satisfy the BAEs (\ref{orBAEs_real}), or
\bea\label{energy_image}
&&\hspace{-0.2truecm}E
\!=\!\frac{N\!\cosh\gamma[\cosh(2\gamma)\!-\!\cos^2(2a)]}{\sinh^2\gamma}\!-\! [\cosh(2\gamma)\!-\!\cos(4 a)] \no \\ &&\!\!\times\!\! \sum^{M}_{j=1}\!\bigg\{\! \frac{1}{\cosh\gamma\!-\!\cos(u_j\!+\!2a)}\!+\!\frac{1}{\cosh\gamma\!-\!\cos(u_j\!-\!2a)} \!\bigg\}\!,
\eea
where $\eta=i\gamma$ is imaginary and $\{u_j\}$ should satisfy the BAEs (\ref{orBAEs_image}).

Next, we check above results numerically.
Numerical solutions of the BAEs and exact diagonalization of the Hamiltonian (\ref{Ham}) are performed for the case of $2N = 4$ and randomly choosing
of model parameters. The results are listed in Table \ref{roots_real} for real $\eta$ and Table \ref{roots_image} for imaginary $\eta$. We note
that the eigenvalues obtained by solving the BAEs
are exactly the same as those obtained by the exact diagonalization of the Hamiltonian (\ref{Ham}). The
energies of the system are degenerated and there are only 8 separated energy level. Therefore, the expression (\ref{energy_real}) or (\ref{energy_image}) gives the complete spectrum of the system.
\begin{table}[!htbp]
\caption{Numerical solutions of the BAEs (\ref{orBAEs_real}) for real $\eta$ case, where $\eta=1$, $b=1$, $2N=4$, $n$ indicates the number of the energy levels
and $E_n$ is the corresponding energy.
The energy $E_n$ calculated from the Bethe roots is exactly the same as that from the exact diagonalization of the Hamiltonian (\ref{Ham}). }\label{roots_real}
\begin{tabular}{|c|c|c|c|}
\hline $ u_1 $ & $ u_2 $ & $ E_n $ & $ n $\\ \hline
$-2.0080+0.0000i$ & $2.0080+0.0000i$ & $-100.4304$ &  $1$ \\ \hline
 $2.6286+0.0000i$ & $---$ & $-20.0748$ &  $2$ \\ \hline
 $-2.0253-3.1416i$ & $2.0253+0.0000i$ & $-20.0748$ &  $2$ \\ \hline
 $0.0000+0.0000i$ & $---$ & $5.0260$ &  $3$ \\ \hline
 $0.0000-3.1416i$ & $0.0000+0.0000i$ & $17.9135$ &  $4$ \\ \hline
 $0.0000-1.3032i$ & $0.0000+1.3032i$ & $18.1853$ &  $5$ \\ \hline
 $---$ & $---$ & $22.2360$ &  $6$ \\ \hline
 $0.0000-3.1416i$ & $---$ & $35.1235$ &  $7$ \\ \hline
 $-2.0777-3.1416i$ & $2.0777-3.1416i$ & $60.0091$ &  $8$ \\
\hline\end{tabular}
 \end{table}
\begin{table}[!htbp]
\caption{ Numerical solutions of the BAEs (\ref{orBAEs_image}) for imaginary $\eta$ case, where $\gamma=1$, $a=1$ and $2N=4$. }\label{roots_image}
\begin{tabular}{|c|c|c|c|}
\hline $ u_1 $ & $ u_2 $ & $ E_n $ & $ n $\\ \hline
$-1.9566+0.0000i$ & $1.9566+0.0000i$ & $-12.1765$ &  $1$ \\ \hline
 $-3.1416+0.0000i$ & $0.0000+0.0000i$ & $-4.3247$ &  $2$ \\ \hline
 $-1.8439+0.0000i$ & $---$ & $-1.8476$ &  $3$ \\ \hline
 $-1.5708-0.9497i$ & $-1.5708+0.9497i$ & $-1.8476$ &  $3$ \\ \hline
 $-3.1416+0.0000i$ & $---$ & $0.1830$ &  $4$ \\ \hline
 $-3.1416-1.1002i$ & $-3.1416+1.1002i$ & $1.1932$ &  $5$ \\ \hline
 $0.0000-1.3426i$ & $0.0000+1.3426i$ & $2.9633$ &  $6$ \\ \hline
 $0.0000+0.0000i$ & $---$ & $3.5122$ &  $7$ \\ \hline
 $---$ & $---$ & $8.0199$ &  $8$ \\
 \hline\end{tabular}
\end{table}

\section{Ground state and elementary excitations for real $\eta$}\label{4}

In this section we study the ground state and elementary excitations of the system.
First, we consider the real $\eta$ case. Taking the logarithm of BAEs (\ref{orBAEs_real}), we have
\begin{eqnarray}\label{logBAEs}
&&  N [\theta_1(u_j\!+\!2b) \!+\! \theta_1(u_j\!-\!2b)]=2 \pi I_j \!+\!\sum^M_{k=1} \theta_2(u_j\!-\!u_k), \no \\ && \quad \quad j=1,\ldots,M,
\end{eqnarray}
where
\begin{equation}\label{theta}
  \theta_n(x)=2\arctan \frac{\tanh (x/2)}{\tan (n\eta/2)}.
\end{equation}
Here the quantum number $\{ I_j \}$ are certain integers (or half odd integers) if $M$ is odd (or even).
For convenience, we define the counting function
\begin{equation}\label{cou-fun}
 Z(u)\!=\!\frac{1}{4\pi} \! \bigg[   \theta_1(u+2b) \!+\! \theta_1(u-2b) \!-\! \frac{1}{N}\!\sum^M_{k=1} \theta_2(u\!-\!u_k) \bigg]\! .\!\!
\end{equation}
Obviously, $Z(u_j ) = I_j/2N$ corresponds to the Eq.(\ref{logBAEs}).
In the thermodynamic limit, $N \rightarrow \infty$, $M \rightarrow \infty$ and $N /M$  finite, taking the derivative of Eq.(\ref{cou-fun}) with respect to $u$, we obtain
\begin{eqnarray}\label{den-1}
\frac{d Z(u)}{d u} &=& \frac{1}{2}  \left[ a_1(u+2b) +a_1(u-2b) \right] - \int^\infty_{-\infty} a_2(u-\lambda) \no \\ && \times \rho(\lambda) d\lambda \nonumber  \\
&\equiv &  \rho(u)+\rho^h(u),
\end{eqnarray}
where
\begin{equation}\label{afun}
 a_n(x)= \frac{1}{2\pi}\frac{\sin(n\eta)}{\cosh u-\cos(n\eta)},
\end{equation}
$\rho(x)$ and $\rho^h(x)$ are the densities of particles and holes, respectively.

\subsection{Ground state}

\begin{figure}[htbp]
\centering
\includegraphics[height=4cm,width=7cm]{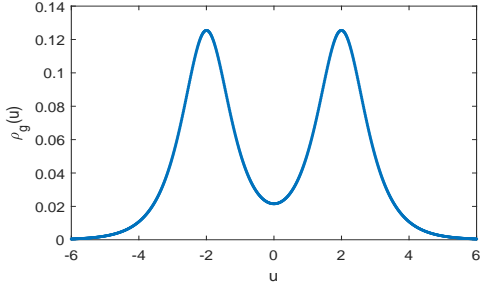}
\caption{The density of Bethe roots at the ground state with $\eta=1$ and $b=1$.}\label{GS-density}
\end{figure}

From the analysis of Eq.(\ref{logBAEs}), we know that at the ground state, $M=N$ which is half of the number of sites.
Meanwhile, all the Bethe roots constrained by Eq.(\ref{logBAEs}) are real and the corresponding quantum numbers are
\begin{equation}\label{GSqnumber}
  I_j=-\frac{N-1}{2}, -\frac{N-3}{2}, \cdots , \frac{N-1}{2}.
\end{equation}
From Eq.(\ref{energy_real}), we learn that each real Bethe root $u_j$ contributes a negative
energy. At the ground state, the Bethe roots should fill the whole real axis and leave
no hole, i.e., $\rho^h(u) = 0$. This means that the density of particles $\rho_g(u)$ at
the ground state satisfies
\begin{eqnarray}\label{den-2}
\rho_g(u) &=& \frac{1}{2}  \left[ a_1(u+2b) +a_1(u-2b) \right] -  \int^\infty_{-\infty} a_2(u-\lambda) \no \\ && \times \rho_g(\lambda) d\lambda.
\end{eqnarray}
Let us define the following Fourier transformation
\begin{eqnarray}
  \tilde{f}(\o ) &=& \int^\infty_{-\infty} f(u) e^{i\o u} du, \no \\
  f(u) &=& \frac{1}{2\pi}\int^\infty_{-\infty} \tilde{f}(\o ) e^{-i\o u} d\o.
\end{eqnarray}
Without losing generality, we consider the case $\eta\in(0,\pi)$. Taking the Fourier transformation of Eq.(\ref{den-2}), we obtain
\begin{eqnarray}
&& \tilde{a}_n(\o ) = \frac{\sinh(\pi\o-2\delta_n\pi\o)}{\sinh\pi\o}. \no \\
&& \tilde{\rho}_g(\o ) = \frac{\cos(2b\, \o )}{2\cosh(\eta\,\o)},
\end{eqnarray}
with $\delta_n\equiv\frac{n\eta}{2\pi}-\lfloor\frac{n\eta}{2\pi}\rfloor$ denoting the fractional part of $\frac{n\eta}{2\pi}$.
Thus the solution of Eq.(\ref{den-2}) is
\begin{equation}
  \rho_g(u) = \frac{1}{8\eta}\left[ \frac{1}{\cosh(\frac{\pi(u+2b)}{2\eta})}+\frac{1}{\cosh(\frac{\pi(u-2b)}{2\eta})} \right].
\end{equation}
The Bethe root distribution at the ground state is shown in Fig.\ref{GS-density}.
The magnetization at the ground state is
\begin{equation}
 \frac12-\int^\infty_{-\infty} \rho_g(u) du = 0,
\end{equation}
indicating a singlet ground state.
The energy density at the ground state reads
\begin{eqnarray}
e_g
&=&\frac{\!\cos\eta[\cosh^2(2b)\!-\!\cos(2\eta)]}{2\sin^2\eta}\!-\! \frac{\cosh(4b)-\cos(2\eta)}{\sin\eta}
\no \\ &\times&\!\!2\pi \int^\infty_{-\infty} [ a_1(u+2b)+a_1(u-2b) ] \rho_g(u) du \no \\
&=&\frac{\!\cos\eta[\cosh^2(2b)\!-\!\cos(2\eta)]}{2\sin^2\eta}\!-\! \frac{\cosh(4b)-\cos(2\eta)} {\sin\eta}
\no \\ &\times&\!\! \int^\infty_{-\infty} \frac{\sinh(\pi \o-\eta \o) \cos^2(2b \o)}{\sinh(\pi \o) \cosh(\eta \o)} d\o.
\end{eqnarray}

\subsection{Spinon Excitations}

\begin{figure*}[htbp]
    \begin{minipage}[b]{0.4\textwidth}
      \centering
      \includegraphics[width=1\textwidth]{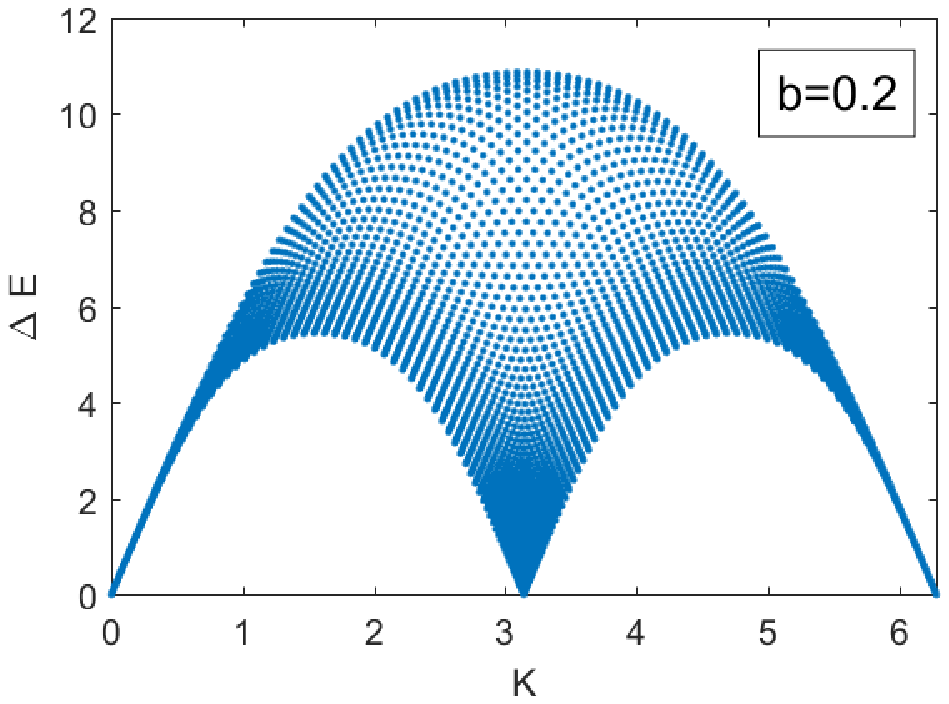}
    \end{minipage}
    \mbox{\hspace{0.80cm}}
    \begin{minipage}[b]{0.4\textwidth}
      \centering
      \includegraphics[width=1\textwidth]{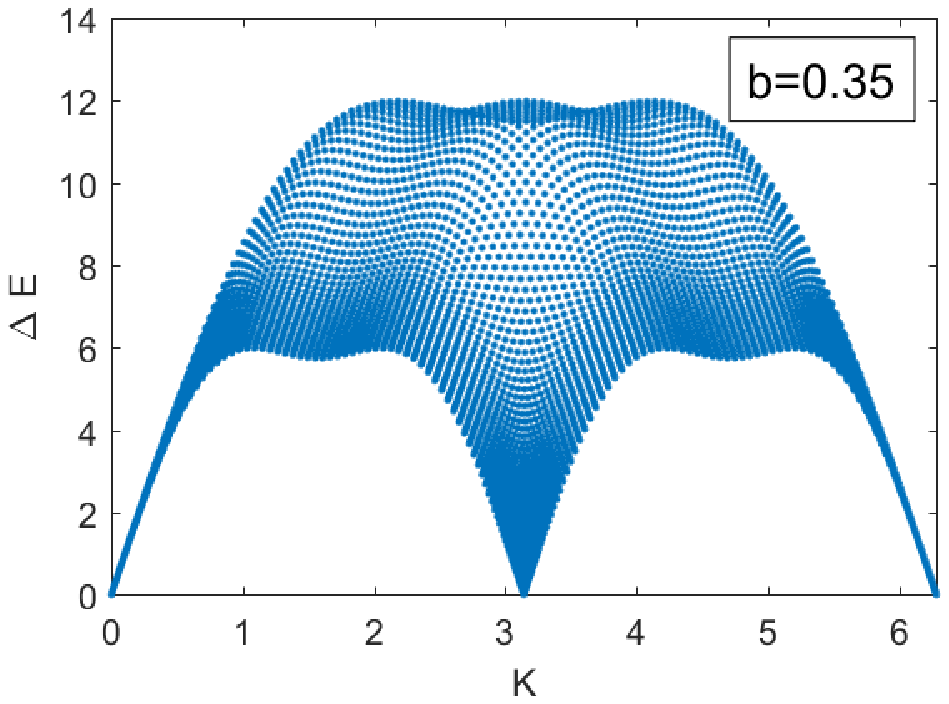}
    \end{minipage}
    \mbox{\hspace{-0.10cm}}
    \begin{minipage}[b]{0.4\textwidth}
      \centering
      \includegraphics[width=1\textwidth]{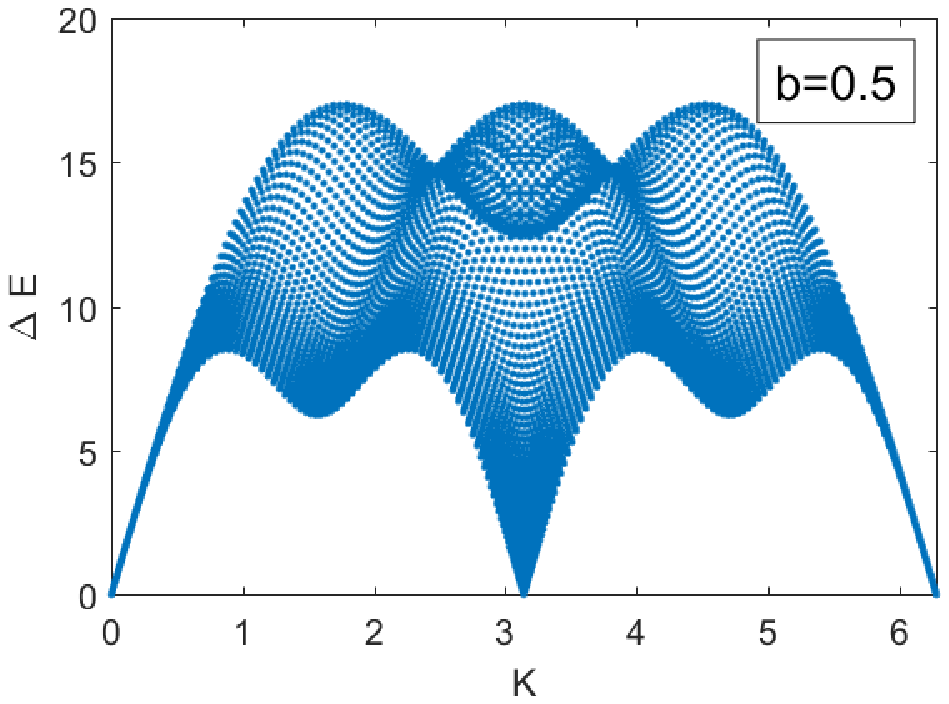}
    \end{minipage}
    \mbox{\hspace{0.70cm}}
    \begin{minipage}[b]{0.4\textwidth}
      \centering
      \includegraphics[width=1\textwidth]{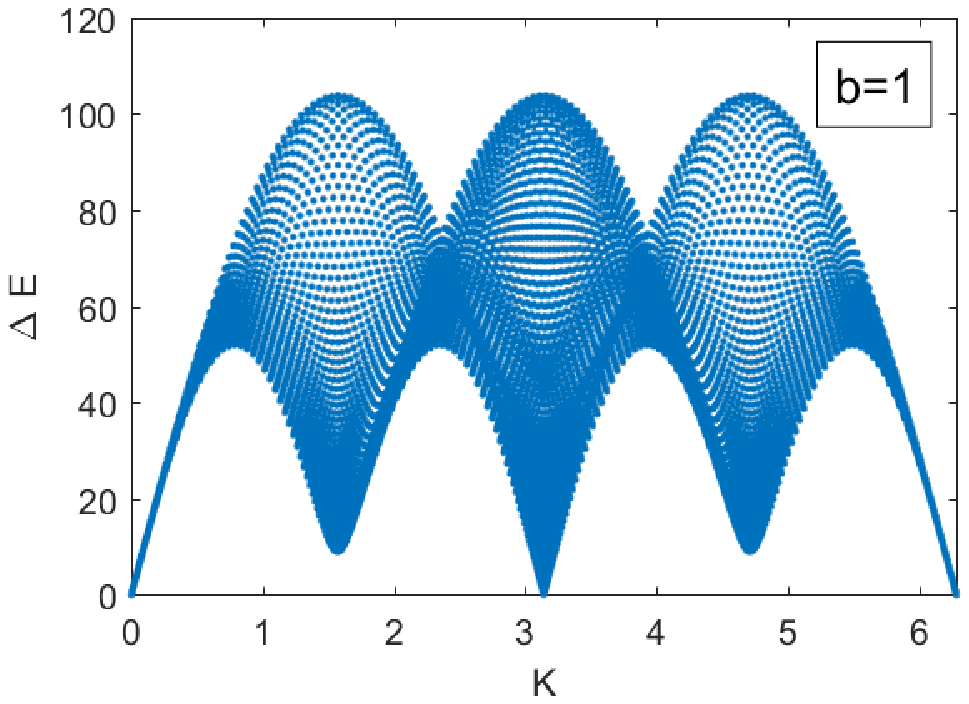}
    \end{minipage}
    \caption{The dispersion relations between energy $\Delta E$ and momentum $K$ of spinon excitations with $\eta=1$.}\label{EEspec}
\end{figure*}

Now we consider the elementary excitations. The simplest excitation is the case of one less spin flipped, i.e., $M = N-1$.
Such a configuration is described by putting two holes in the Fermi sea. Meanwhile, all the Bethe roots constrained by Eq.(\ref{logBAEs}) are real and the corresponding quantum numbers are
\begin{eqnarray}
  I_j&=&-\frac{N}{2},-\frac{N}{2}+1,\cdots, -\frac{N}{2}+r-1,-\frac{N}{2}+r+1, \cdots, \no \\
  && -\frac{N}{2}+s-1,-\frac{N}{2}+s+1,\cdots, \frac{N}{2}-1, \frac{N}{2},\label{EEqnumber}
\end{eqnarray}
where $0 < r < s < N$. The positions of holes are denoted as $u^h_r$
and $u^h_s$. In this case all $N - 1$ quasi-momentum are real numbers and
the total momentum is $\Delta K = \pi(r + s)/N$. In the thermodynamic limit, the momentum of this excitation is calculated as
\begin{eqnarray}\label{K-h}
\hspace{-1.2truecm} K &=& 2\pi \int^\infty_{u^h_r} \rho_g(u) du + 2\pi \int^\infty_{u^h_s} \rho_g(u) du \no \\
\hspace{-0.28truecm}&=&\arctan[e^{-\frac{\pi(u^h_r-2b)}{2\eta}}] +\arctan[e^{-\frac{\pi(u^h_r+2b)}{2\eta}}]
\no \\ && +\arctan[e^{-\frac{\pi(u^h_s-2b)}{2\eta}}]+\arctan[e^{-\frac{\pi(u^h_s+2b)}{2\eta}}].
\end{eqnarray}
The density of holes is
\begin{eqnarray}\label{den_h}
\rho^h(u) = \frac{1}{2N}  \left[ \delta_1(u-u^h_r) + \delta_1(u-u^h_s) \right].
\end{eqnarray}
The corresponding Bethe root density becomes $\rho_e(u)=\rho_g(u)+\delta\rho(u)$ \cite{Tak99}. The density $\rho_e(u)$ will deviate from $\rho_g(u)$ by $\delta \rho(u)$ because of the presence of the two holes. From Eqs.(\ref{den-1}) and (\ref{den_h}), we obtain
\begin{eqnarray}\label{den-h1}
\rho_e(u)+\rho^h(u) &=& \frac{1}{2}  \left[ a_1(u+2b) +a_1(u-2b) \right] \no \\ && -  \int^\infty_{-\infty} a_2(u-\lambda) \rho_e(\lambda) d\lambda.
\end{eqnarray}
After some calculations, we have
\begin{eqnarray}\label{den-h2}
\delta\tilde{\rho}(\omega) = -\frac{1}{2N}  \frac{e^{i\omega u^h_r}+e^{i\omega u^h_s}}{1+\tilde{a}_2(\o)}.
\end{eqnarray}
The excitation energy is
\begin{eqnarray}\label{E-h}
\Delta E &=& -\frac{4\pi N [\cosh(4b)-\cos(2\eta)]}{\sin\eta} \int^\infty_{-\infty} [a_1(u+2b) \no \\ && + a_1(u-2b)] \delta\rho(u) du \no \\
&=&\frac{4\pi [\cosh(4b)-\cos(2\eta)]}{\sin\eta} [\rho_g(u^h_r)+\rho_g(u^h_s)] \no \\
&=&\varepsilon(u^h_r)+\varepsilon(u^h_s),
\end{eqnarray}
where
\begin{eqnarray}\label{e-h}
\varepsilon(u) = \frac{4\pi [\cosh(4b)-\cos(2\eta)]}{\sin\eta} \rho_g(u).
\end{eqnarray}
We see that the energy of such an excitation is the summation of the energies of two holes.
Here the two holes together carry spin-1, and each of them carries spin-$\frac{1}{2}$. These excitations
are usually called spinons \cite{Fad81}, a typical fractional excitation in the one-dimensional
quantum systems.

The dispersion relation of the spinon excitations can be derived from equations (\ref{K-h}) and (\ref{E-h}). The numerical results are shown in Fig.\ref{EEspec}. From it, we see that the spinon excitation is gapless
which can be reached by putting the holes at the points $0$ or $\pm\pi$. Meanwhile, if $b$ is very small, the excitation spectrum is quite similar to that of the conventional XXZ model \cite{Tak99}. With the increasing of $b$, the excitation spectrum turns to the triple arched structure. Unlike the conventional $J_1-J_2$ model, there is no dimerization in the present model for any real $b$. If $b$ takes some imaginary values, dimerization indeed occurs as hinted from the solution of the BAE's. However, in such a case the Hamiltonian is non-hermitian.

\section{Ground state and elementary excitations for imaginary $\eta$}\label{5}

\begin{figure}[htbp]
\centering
\includegraphics[height=4cm,width=7cm]{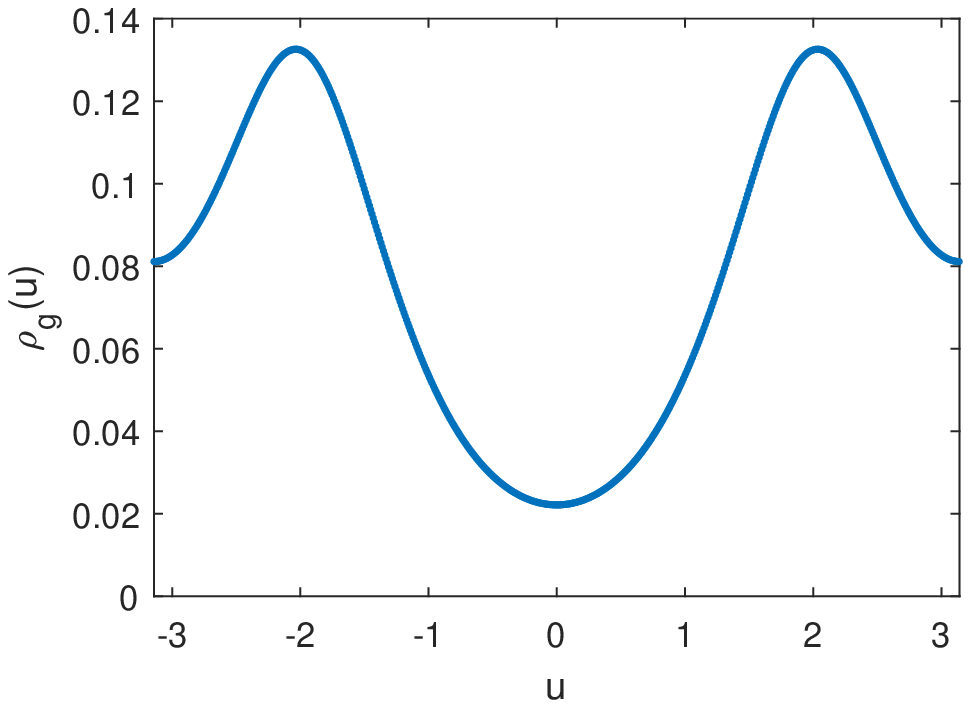}\\
\includegraphics[height=4cm,width=7cm]{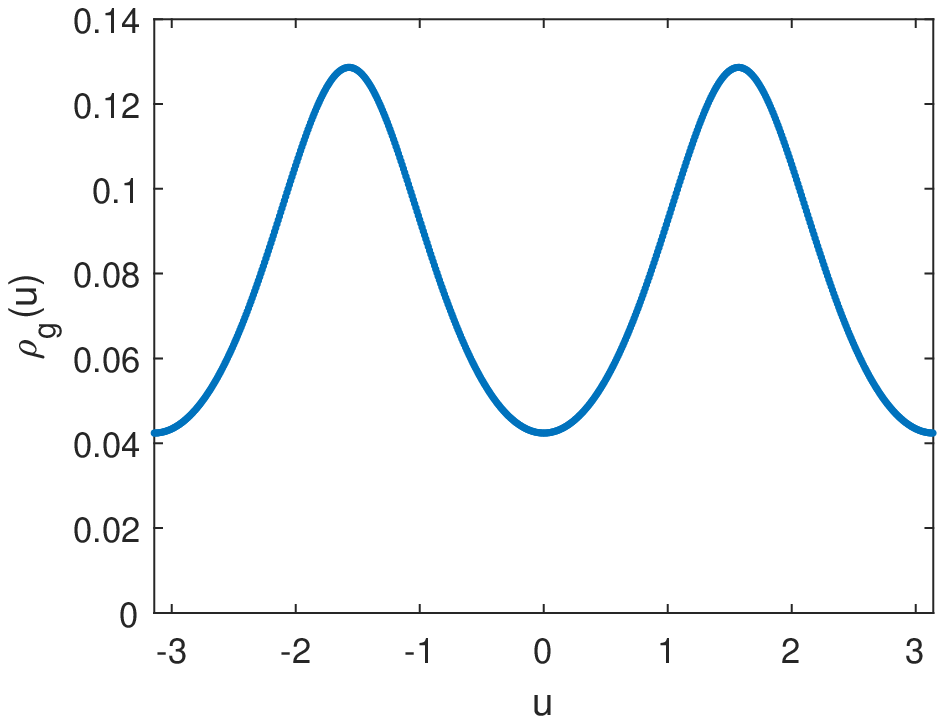}
\caption{The densities of Bethe roots at the ground state with $\gamma=1$, $a=1$ (upper); and $\gamma=1$, $a=\pi/4$ (lower).}\label{GS-density-image}
\end{figure}

In this section we study the ground state and elementary excitations of the system for imaginary $\eta=i\gamma$.
Without losing generality, we assume $\gamma > 0$. Similarly, let us introduce
\begin{equation}\label{afun-image}
 a_n(x)= \frac{1}{2\pi}\frac{\sinh(n\gamma)}{\cosh (n\gamma)-\cos x}.
\end{equation}
The Fourier transformation of $a_n(x)$ is
\begin{equation}\label{afun-FT-image}
  \tilde{a}_n(\o)\!=\!\frac{1}{2\pi}\int_{-\pi}^{\pi}\!e^{i\o x}\frac{\sinh(n\gamma)}{\cosh (n\gamma)-\cos x}dx\!=\!e^{-n\gamma|\o|}.\!
\end{equation}
Note that $\o$ takes values of integers. From the energy expression (\ref{energy_image}), we know that at the ground state the Bethe roots still take
real values and fill the region $(-\pi, \pi]$. Using the similar procedure mentioned above, we find the density of Bethe roots at the ground state satisfies
\begin{eqnarray}\label{den-2-image}
\rho_g(u) &=& \frac{1}{2}  \left[ a_1(u+2a) +a_1(u-2a) \right] -  \int^\pi_{-\pi} a_2(u-\lambda) \no \\ && \times \rho_g(\lambda) d\lambda.
\end{eqnarray}
Using Fourier transformation we have
\begin{eqnarray}
&&\tilde{\rho}_g(\o ) = \frac{\cos(2a\, \o )}{2\cosh(\gamma\,\o)}, \no \\
&&\rho_g(u) = \frac{1}{2\pi}\sum_{\o=-\infty}^{\infty}e^{-iu\o} \frac{\cos(2a\,\o)}{2\cosh(\gamma \o)}.
\end{eqnarray}
The Bethe root distribution at the ground state is shown in Fig.\ref{GS-density-image}.
Interestingly, we find $\rho_g(\pm\pi)=\rho_g(0)$ if $a=\pi/4+k\pi/2$, and $k$ is an arbitrary integer. Please see the lower one in Fig.\ref{GS-density-image}.
The total magnetization at the ground state is still zero.
The energy density at the ground state reads
\begin{eqnarray}
  e_g &=& \frac{\!\cosh\gamma[\cosh(2\gamma)\!-\!\cos^2(2a)]}{2\sinh^2\gamma}+ \frac{\cos(4a)-\cosh(2\gamma)}{\sinh\gamma} \no \\
  &\times& \sum_{\o=-\infty}^{\infty}\frac{\cos^2(2a\o)e^{-\gamma|\o|}}{\cosh(\gamma\o)}.
\end{eqnarray}

\begin{figure*}
    \begin{minipage}[b]{0.4\textwidth}
      \centering
      \includegraphics[width=1\textwidth]{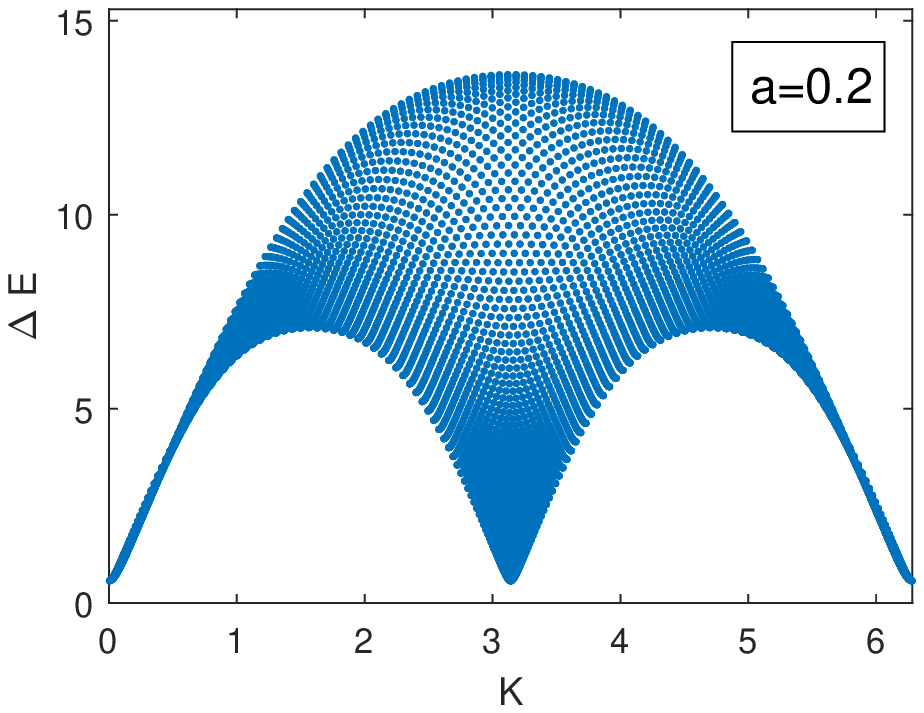}
    \end{minipage}
    \mbox{\hspace{0.80cm}}
    \begin{minipage}[b]{0.4\textwidth}
      \centering
      \includegraphics[width=1\textwidth]{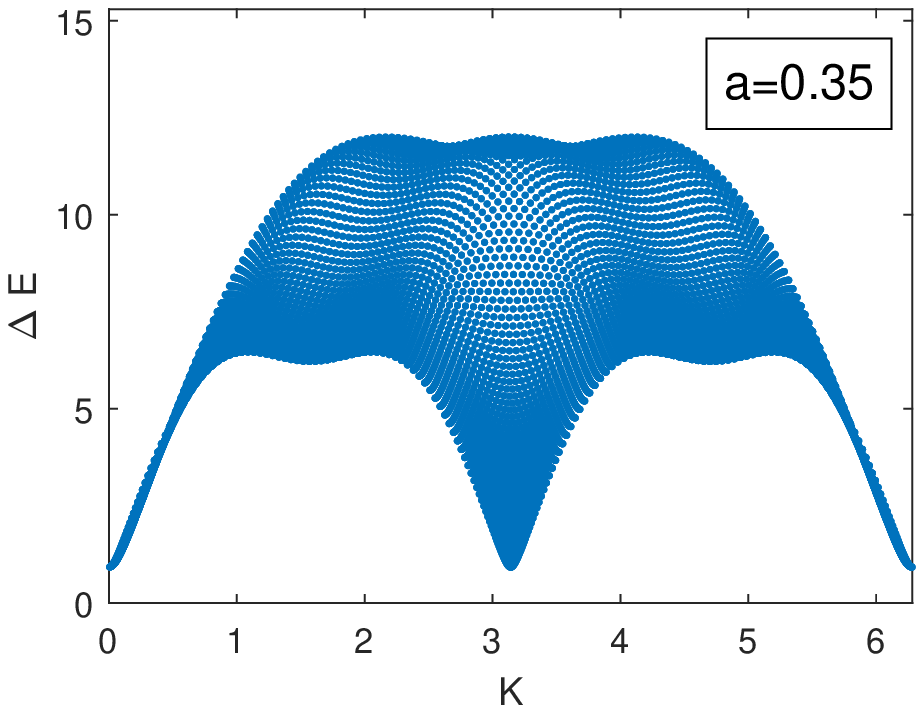}
    \end{minipage}
    \mbox{\hspace{-0.10cm}}
    \begin{minipage}[b]{0.4\textwidth}
      \centering
      \includegraphics[width=1\textwidth]{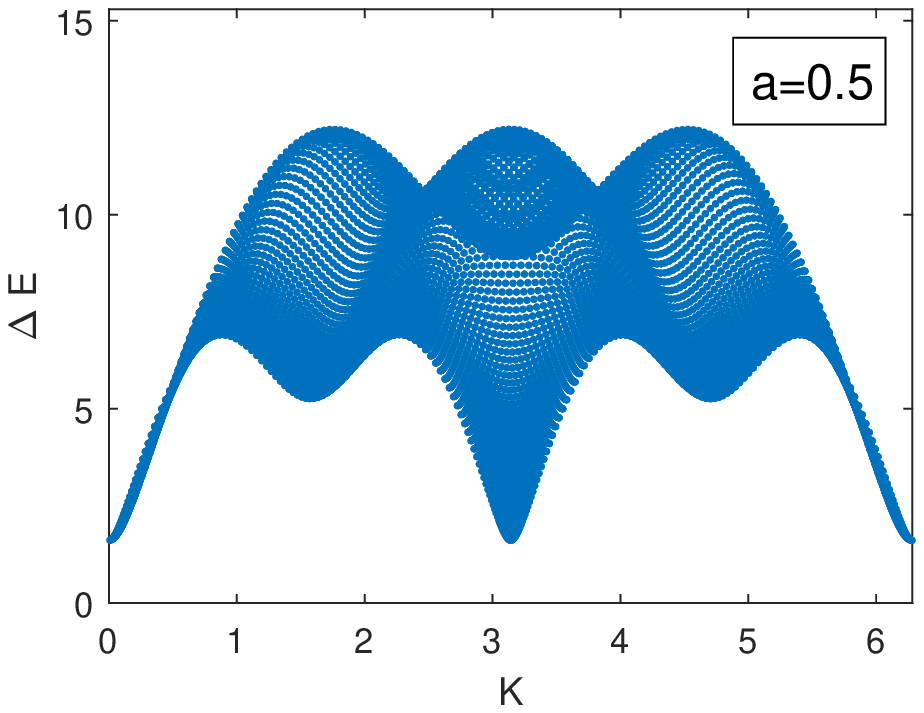}
    \end{minipage}
    \mbox{\hspace{0.70cm}}
    \begin{minipage}[b]{0.4\textwidth}
      \centering
      \includegraphics[width=1\textwidth]{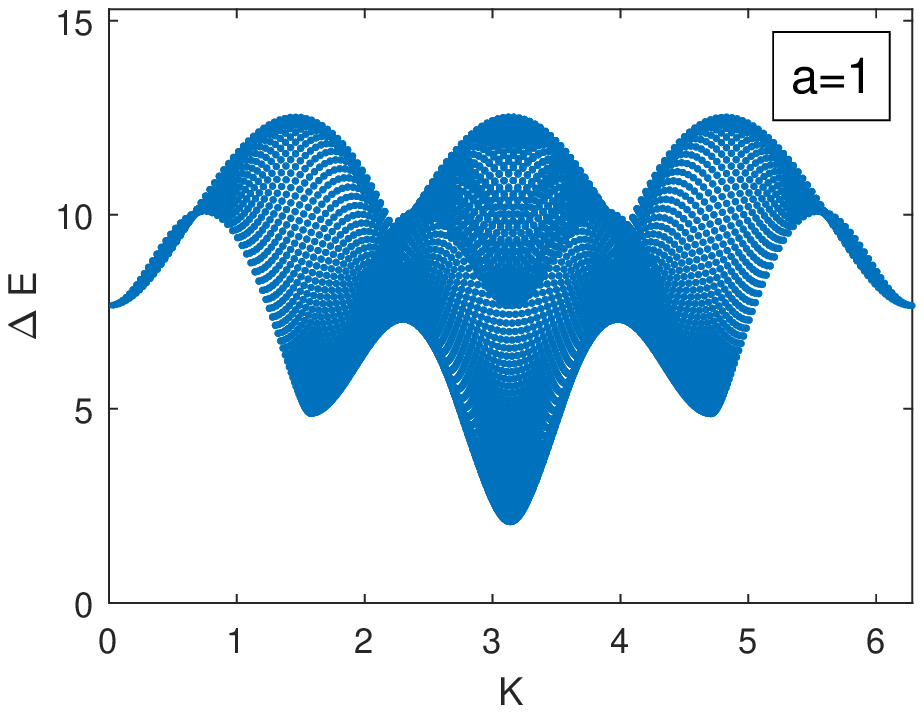}
    \end{minipage}
    \caption{The dispersion relations between energy $\Delta E$ and momentum $K$ of spinon excitations with $\gamma=1$.}\label{EEspec-image}
\end{figure*}

In the thermodynamic limit, the momentum of spinion excitation is calculated as
\begin{eqnarray}\label{K-h-image}
\hspace{-0.5truecm} K &=& 2\pi \int^\pi_{u^h_r} \rho_g(u) du + 2\pi \int^\pi_{u^h_s} \rho_g(u) du
\no \\ \hspace{-0.28truecm}&=&\sum_{\substack{\o=-\infty,\\ \o\neq 0}}^{\infty} \!\!\frac{\cos(2a\o)}{2i\o\cosh(\gamma\o)} [2(-1)^\o-e^{i\o u_r^h}-e^{i\o u_s^h}] \no \\
&&+\pi-\frac{u_r^h+u_s^h}{2}.
\end{eqnarray}
After some calculations similar with above real $\eta$ case, we obtain the excitation energy
\begin{eqnarray}\label{E-h-image}
\Delta E=\varepsilon(u^h_r)+\varepsilon(u^h_s),
\end{eqnarray}
where
\begin{eqnarray}\label{e-h-image}
\varepsilon(u) = \frac{4\pi [\cosh(2\gamma)-\cos(4a)]}{\sinh\gamma} \rho_g(u).
\end{eqnarray}

The dispersion relation of the spinon excitations can be derived from equations (\ref{K-h-image}) and (\ref{E-h-image}). The numerical results are shown in Fig.\ref{EEspec-image}. From it, we see that if $a$ is very small, the excitation spectrum is quite similar to that of the conventional XXZ model \cite{Tak99}. With the increasing of $a$, the excitation spectrum turns to the triple arched structure. Similar with real $\eta$ case, there is also no dimerization in the present model for any imaginary $\eta$ and real $a$.

From the excitation spectrum in Fig.\ref{EEspec-image}, we also find that the elementary excitations possess a finite gap.
Now we determine the values of the gap. Without losing generality, we assume $a\in[0,\pi]$. We should consider the positions of holes first.
From Fig.\ref{GS-density-image}, we know that there are two minimal points located at $u=0$ and $u=\pi$.
As we mentioned before, $\rho_g(\pi)=\rho_g(0)$ if $a=\pi/4$ or $a=3\pi/4$. By detailed analysis, we conclude that
$\rho_g(\pi)<\rho_g(0)$ if $0<a<\pi/4$ or $3\pi/4<a<\pi$ and $\rho_g(\pi)>\rho_g(0)$ if $\pi/4<a<3\pi/4$.
Thus we put the holes at the point of $\pi$ if $a\in[0,\pi/4]\cup[3\pi/4,\pi]$, and put the holes at the point of $0$ if $a\in(\pi/4,3\pi/4)$.
We note that only in the thermodynamic limit, two holes can be put at the same position.
After some calculations, we obtain the energy gap of the model (\ref{Ham}) in these intervals as
\begin{equation}\label{gap-1}
  \Delta = \frac{4[\cosh(2\gamma)\!-\!\cos(4a)]}{\sinh\gamma}\!\sum_{\o=-\infty}^{\infty}\!\! \frac{(-1)^\o\!\cos(2a\o)}{2\cosh(\gamma\o)},
\end{equation}
if $a\in[0,\pi/4]\cup[3\pi/4,\pi]$, and
\begin{equation}\label{gap-2}
  \Delta = \frac{4[\cosh(2\gamma)\!-\!\cos(4a)]}{\sinh\gamma}\sum_{\o=-\infty}^{\infty} \frac{\cos(2a\o)}{2\cosh(\gamma\o)},
\end{equation}
if $a\in(\pi/4,3\pi/4)$.

\begin{figure}[htbp]
\centering
\includegraphics[height=4cm,width=7cm]{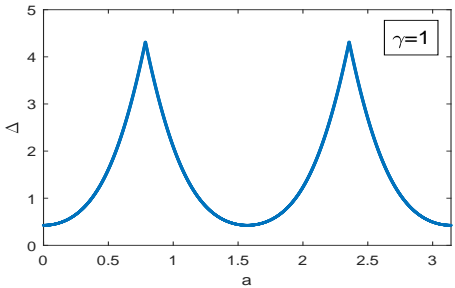}
\caption{The gap with $\gamma=1$.}\label{gap}
\end{figure}

The gap is shown in Fig.\ref{gap}. From it, we see that the gap is enhanced by the NNN and chiral three-spin interactions. At the points of $a=0$ and $a=\pi$,
the gap takes its minimum, which is the same as that of the conventional XXZ spin chain, because that the model (\ref{Ham}) degenerates into the conventional XXZ spin chain at these points.
At the point of $a=\frac{\pi}2$, the gap also takes its minimum. In this case the model (\ref{Ham}) degenerates into the XXZ spin chain only with NN interaction where the couplings along the $x$ and $y$ directions are negative.
At the points of $a=\frac{\pi}{4}$ and $a=\frac{3\pi}{4}$, the gap arrives at its maximal value.
This is because at the these points, the coupling strengths of NNN and chiral three-spin interactions
reach their maximum and the NN couplings along the $x$ and $y$ directions are zero.
Besides, the gap also has the property
\begin{eqnarray}
\Delta(a) =\Delta(\frac{\pi}{2}-a) =\Delta(\frac{\pi}{2}+a) =\Delta(\pi-a),\no
\end{eqnarray}
which means that the gap is symmetric with respect to the points of $\frac{\pi}{4}$, $\frac{\pi}{2}$ and $\frac{3\pi}{4}$.
This symmetry is different from that of the Hamiltonian, where the Hamiltonian is symmetric only with respect to $\pi$, i.e.,
\begin{eqnarray}
 H(a)=H(\pi+a).\no
\end{eqnarray}

\section{Non-hermitian case}\label{6}

In this section, we consider the case that both $a$ and $\eta$ are real, which implies that the Hamiltonian (\ref{Ham}) is non-hermitian.
By the analysis of possible couplings in the Hamiltonian (\ref{Ham}), we restrict the values of $a$ in the interval $[0,\pi]$.
It is easy to check that following identity holds
\begin{equation}\label{symm}
  H(a,\pi+\eta)=H(\pi-a,\pi-\eta).
\end{equation}
This means that the values of $\eta$ can also be restricted in the interval $(0,\pi)$. Then the parameters $a\in [0,\pi]$ and $\eta\in (0,\pi)$ can describe all the
coupling strengths.

Using the direct diagonalization method, up to $2N=12$, we find that all the eigenvalues of the Hamiltonian (\ref{Ham}) are real if $a$ takes the values in some intervals.
After detailed calculation, the intervals are determined as $a\in[0,\eta/2]\cup[(\pi-\eta)/2,(\pi+\eta)/2]\cup[\pi-\eta/2,\pi]$. We note that if $\eta>\pi/2$,
these intervals are connected with each other and the parameter $a$ fills the whole interval $[0,\pi]$, which means that with an arbitrary $a$,
all the eigenvalues of the model (\ref{Ham}) are real.

The BAEs (\ref{orBAEs1}) are true for real $a$ and real $\eta$. Put $\lambda_j = i u_j/2 - \eta/2$ and the BAEs become
\bea\label{orBAE1s}
&&\left[\frac{\sinh\frac{1}{2}(u_j-i(2a+\eta))\sinh\frac{1}{2}(u_j+i(2a-\eta))}
{\sinh\frac{1}{2}(u_j+i(2a+\eta))\sinh\frac{1}{2}(u_j-i(2a-\eta))}\right]^N
\no \\ && \quad =\prod^{M}_{l\neq j}\frac{\sinh\frac{1}{2}(u_j-u_l-2\eta i)}{\sinh\frac{1}{2}(u_j-u_l+2\eta i)}, \quad j=1,\cdots,M.
\eea
From Eqs.(\ref{Hdef}) and (\ref{Lam}) we obtain the eigenvalue of the Hamiltonian (\ref{Ham}) in terms of the Bethe roots as
\bea\label{energ1y}
&&\hspace{-0.5truecm}E
\!=\!\frac{N\!\cos\eta[\cos^2(2a)\!-\!\cos(2\eta)]}{\sin^2\eta}\!-\! [\cos(4 a)\!-\!\cos(2\eta)]
\no \\ &&\!\!\!\!\times\!\! \sum^{M}_{j=1}\!\bigg\{\! \!\frac{1}{\cosh(u_j\!+\!2ai)\!-\!\cos\!\eta}\!+\!\frac{1}{\cosh(u_j\!-\!2ai)\!-\!\cos\!\eta} \!\!\bigg\}\!,
\eea
Solving the BAEs (\ref{orBAE1s}) with $2N=6$ and substituting the values of Bethe roots into (\ref{energ1y}), we find that the
eigenvalues calculated from the Bethe roots are exactly the same as those obtained from the exact diagonalization of the Hamiltonian (\ref{Ham}). The energy spectrum is complete.
Meanwhile, the intervals that all the eigenvalues are real keep unchanged.

\section{Conclusion}\label{7}

In this paper, we propose a new integrable anisotropic $J_1-J_2$ spin chain model with extra scalar chirality terms.
By means of the Bethe Ansatz method, we obtain the exact solution of the system. The ground state and the
novel structure of the elementary excitation spectrum are obtained. We find that the elementary excitation is gapless if the anisotropic parameter is real
while the elementary excitation has a gap if the anisotropic parameter is imaginary. Moreover, it is shown that the spinon excitation spectrum of the  model possesses a novel triple arched structure. The method of this paper can be used to construct other new integrable models with next-nearest-neighbour couplings.

\section*{Acknowledgments}

We would like to thank Prof. Y. Wang for his valuable discussions and continuous encouragements.
The financial supports from the National Program
for Basic Research of MOST (Grant Nos. 2016YFA0300600 and
2016YFA0302104), the National Natural Science Foundation of China
(Grant Nos. 11434013, 11425522, 11547045, 11774397, 11775178 and 11775177), the Major Basic Research Program of Natural Science of Shaanxi Province
(Grant Nos. 2017KCT-12, 2017ZDJC-32), Australian Research Council (Grant No. DP 190101529), the Strategic Priority Research Program of the Chinese
Academy of Sciences and the Double First-Class University Construction Project of Northwest University are gratefully acknowledged.

\section*{Appendix: Derivation of the Hamiltonian}
\renewcommand{\baselinestretch}{1.6}
\setcounter{equation}{0}
\renewcommand{\theequation}{A\arabic{equation}}

Using the initial condition (\ref{PT}) of the $R$-matrix $R(u)$ given by (\ref{R-matrix}),  we can evaluate the values of
the transfer matrix $\hat t(u)$ at some points: $u=\pm a$:
\bea
&&\hat{t}(-a) =      R_{2N,2N-1}(-2a) \cdots  R_{2N,2}(0) \no \\
&&\qquad\qquad \times R_{2N,1}(-2a),\label{t1} \\[4pt]
&&\hat{t}(a) =      R_{1,2N}(2a) R_{1,2N-1}(0) \cdots  R_{1,2}(2a).\no
\eea
Taking the derivative of the transfer matrix $t(u)$ with respect to $u$ at the point of $u=a$, we have
\bea\label{t3}
&&\frac{\partial \, t(u)}{\partial u}\big|_{u=a}
= \sum^{N-1}_{j=1} R_{2N,1}(2a) R_{2N,2}(0) \cdots
\no \\ && \quad \times [ R'_{2N,2j-1}(2a) R_{2N,2j}(0) + R_{2N,2j-1}(2a)
\no \\[4pt] && \quad \times R'_{2N,2j}(0) ] \cdots R_{2N,2N-1}(2a)
+ R_{2N,1}(2a) \no \\[4pt] && \quad \times R_{2N,2}(0) \cdots R'_{2N,2N-1}(2a) + R_{2,3}(2a)
\no \\[4pt] && \quad \times R_{2,4}(0) \cdots R_{2,2N-1}(2a) R'_{2,2N}(0)R_{2,1}(2a),
\eea
where $R'_{i,j}(u)=\frac{\partial }{\partial u}\,R_{i,j}(u)$. Similarly we can calculate the derivative of $t(u)$ at the point of $u=-a$
\bea\label{t4}
&& \frac{\partial \, t(u)}{\partial u}\big|_{u=-a} = \sum^{N}_{j=1} R_{1,2}(-2a) \cdots [ R'_{1,2j-1}(0)
\no \\[4pt] && \quad \times R_{1,2j}(-2a)
+ R_{1,2j-1}(0) R'_{1,2j}(-2a) ]  \cdots R_{1,2N-1}(0)
\no \\[4pt] && \quad \times R_{1,2N}(-2a) + R_{1,2}'(-2a) R_{1,3}(0) \cdots
\no \\[4pt] && \quad \times R_{1,2N}(-2a) + R_{3,4}(-2a)R_{3,5}(0) \cdots
\no \\[4pt] && \quad \times R_{3,2N}(-2a) R'_{3,1}(0)R_{3,2}(-2a).
\eea
Substituting the above relations (\ref{t1})-(\ref{t4}) into the expression (\ref{Hdef}), we obtain
\bea\label{Ham-1}
H&=& \sum^{N}_{j=1}
\{  R_{2j,2j-1}(-2a) R'_{2j,2j-1}(2a)
\no \\[4pt] && R_{2j+1,2j}(2a) R'_{2j+1,2j}(-2a) + R_{2j+2,2j+1}(-2a)
\no \\[4pt] && \times P_{2j+2,2j}R'_{2j+2,2j}(0) R_{2j+2,2j+1}(2a)+R_{2j+1,2j}(2a)
\no \\[4pt] && \times P_{2j+1,2j-1} R'_{2j+1,2j-1}(0) R_{2j+1,2j}(-2a) \}
\no \\[4pt] && -\frac{N\cos\eta[\cos^2(2a)-\cos(2\eta)]}{\sin^2\eta}.
\eea
The derivative of the $R$-matrix (\ref{R-matrix}) reads
\bea \label{R'-matrix}
\hspace{-0.6cm} R'_{0,j}(u)\!=\!\frac{1}{2}\!\bigg[ \frac{\cos(u\!+\!\eta)}{\sin\eta}(1 \hspace{-0.08truecm}+\hspace{-0.08truecm}\sigma_0^z\sigma_j^z) \!+\! \frac{\cos u}{\sin\eta}
(1\hspace{-0.08truecm}-\hspace{-0.08truecm}\sigma_0^z\sigma_j^z) \bigg]\!.
\eea
The commutative relation between the permutation operators is
\bea
[P_{2,1},P_{2,0}]
&=&\frac{1}{4}[(1+{\vec\sigma}_{2}\cdot{\vec\sigma}_{1}),(1+{\vec\sigma}_{2}\cdot{\vec\sigma}_{0})] \no \\[4pt]
&=&\frac{i}{2}( \sigma_2^z\sigma_1^x\sigma_0^y - \sigma_2^y\sigma_1^x\sigma_0^z
- \sigma_2^z\sigma_1^y\sigma_0^x
\no \\[4pt] && + \sigma_2^x\sigma_1^y\sigma_0^z + \sigma_2^y\sigma_1^z\sigma_0^x - \sigma_2^x\sigma_1^z\sigma_0^y ) \no \\[4pt]
&=& \frac{i}{2} {\vec\sigma}_2\cdot ({\vec\sigma}_1 \times {\vec\sigma}_0).\label{Rx}
\eea
Substituting Eqs.(\ref{R-matrix}), (\ref{R'-matrix}) and (\ref{Rx}) into (\ref{Ham-1}) and after some tedious calculations, we arrive at the form of the Hamiltonian (\ref{Ham}).

\fussy
\bibliography{ref1}
\end{document}